\documentclass[twocolumn,times]{aastex63}
\usepackage{amsmath}

\setwatermarkfontsize{1.5in}

\newcommand{\insight}{\textit{Insight}-HXMT}

\shortauthors{Zhang ET AL.}
\begin{document}


\title{\insight~observations of the extremely bright GRB 221009A}


\correspondingauthor{Cheng-Kui Li, Shao-Lin Xiong, Shuang-Nan Zhang}
\email{lick@ihep.ac.cn, xiongsl@ihep.ac.cn, zhangsn@ihep.ac.cn}

\author{Wen-Long Zhang}
\affil{Key Laboratory of Particle Astrophysics, Institute of High Energy Physics, Chinese Academy of Sciences, Beijing 100049, China}
\affil{School of Physics and Physical Engineering, Qufu Normal University, Qufu, Shandong 273165, China}

\author{Wang-Chen Xue}
\affil{Key Laboratory of Particle Astrophysics, Institute of High Energy Physics, Chinese Academy of Sciences, Beijing 100049, China}
\affil{University of Chinese Academy of Sciences, Chinese Academy of Sciences, Beijing 100049, China}

\author{Cheng-Kui Li*}
\affil{Key Laboratory of Particle Astrophysics, Institute of High Energy Physics, Chinese Academy of Sciences, Beijing 100049, China}

\author{Shao-Lin Xiong*}
\affil{Key Laboratory of Particle Astrophysics, Institute of High Energy Physics, Chinese Academy of Sciences, Beijing 100049, China}

\author{Gang Li}
\affil{Key Laboratory of Particle Astrophysics, Institute of High Energy Physics, Chinese Academy of Sciences, Beijing 100049, China}

\author{Yong Chen}
\affil{Key Laboratory of Particle Astrophysics, Institute of High Energy Physics, Chinese Academy of Sciences, Beijing 100049, China}

\author{Wei-Wei Cui}
\affil{Key Laboratory of Particle Astrophysics, Institute of High Energy Physics, Chinese Academy of Sciences, Beijing 100049, China}

\author{Xiao-Bo Li}
\affil{Key Laboratory of Particle Astrophysics, Institute of High Energy Physics, Chinese Academy of Sciences, Beijing 100049, China}

\author{Cong-Zhan Liu}
\affil{Key Laboratory of Particle Astrophysics, Institute of High Energy Physics, Chinese Academy of Sciences, Beijing 100049, China}

\author{Ming-Yu Ge}
\affil{Key Laboratory of Particle Astrophysics, Institute of High Energy Physics, Chinese Academy of Sciences, Beijing 100049, China}

\author{Wen-Jun Tan}
\affil{Key Laboratory of Particle Astrophysics, Institute of High Energy Physics, Chinese Academy of Sciences, Beijing 100049, China}
\affil{University of Chinese Academy of Sciences, Chinese Academy of Sciences, Beijing 100049, China}

\author{Jia-Cong Liu}
\affil{Key Laboratory of Particle Astrophysics, Institute of High Energy Physics, Chinese Academy of Sciences, Beijing 100049, China}
\affil{University of Chinese Academy of Sciences, Chinese Academy of Sciences, Beijing 100049, China}

\author{Chen-Wei Wang}
\affil{Key Laboratory of Particle Astrophysics, Institute of High Energy Physics, Chinese Academy of Sciences, Beijing 100049, China}
\affil{University of Chinese Academy of Sciences, Chinese Academy of Sciences, Beijing 100049, China}

\author{Chao Zheng}
\affil{Key Laboratory of Particle Astrophysics, Institute of High Energy Physics, Chinese Academy of Sciences, Beijing 100049, China}
\affil{University of Chinese Academy of Sciences, Chinese Academy of Sciences, Beijing 100049, China}

\author{Yan-Qiu Zhang}
\affil{Key Laboratory of Particle Astrophysics, Institute of High Energy Physics, Chinese Academy of Sciences, Beijing 100049, China}
\affil{University of Chinese Academy of Sciences, Chinese Academy of Sciences, Beijing 100049, China}

\author{Yue Wang}
\affil{Key Laboratory of Particle Astrophysics, Institute of High Energy Physics, Chinese Academy of Sciences, Beijing 100049, China}
\affil{University of Chinese Academy of Sciences, Chinese Academy of Sciences, Beijing 100049, China}

\author{Zhen Zhang}
\affil{Key Laboratory of Particle Astrophysics, Institute of High Energy Physics, Chinese Academy of Sciences, Beijing 100049, China}

\author{Shu-Xu Yi}
\affil{Key Laboratory of Particle Astrophysics, Institute of High Energy Physics, Chinese Academy of Sciences, Beijing 100049, China}

\author{Shuo Xiao}
\affil{School of Physics and Electronic Science, Guizhou Normal University, Guiyang 550001, China}
\affil{Guizhou Provincial Key Laboratory of Radio Astronomy and Data Processing, Guizhou Normal University, \\Guiyang 550001, China}

\author{Ce Cai}
\affil{College of Physics and Hebei Key Laboratory of Photophysics Research and Application, 
\\Hebei Normal University, Shijiazhuang, Hebei 050024, China}

\author{Shuang-Xi Yi}
\affil{School of Physics and Physical Engineering, Qufu Normal University, Qufu, Shandong 273165, China}

\author{Li-Ming Song}
\affil{Key Laboratory of Particle Astrophysics, Institute of High Energy Physics, Chinese Academy of Sciences, Beijing 100049, China}

\author{Lian Tao}
\affil{Key Laboratory of Particle Astrophysics, Institute of High Energy Physics, Chinese Academy of Sciences, Beijing 100049, China}

\author{Shu Zhang}
\affil{Key Laboratory of Particle Astrophysics, Institute of High Energy Physics, Chinese Academy of Sciences, Beijing 100049, China}

\author{Shuang-Nan Zhang*}
\affil{Key Laboratory of Particle Astrophysics, Institute of High Energy Physics, Chinese Academy of Sciences, Beijing 100049, China}
\affil{University of Chinese Academy of Sciences, Chinese Academy of Sciences, Beijing 100049, China}

\begin{abstract} 
The Hard X-ray Modulation Telescope (\insight) detected GRB 221009A, the brightest gamma-ray burst observed to date, with all its three telescopes, i.e. High Energy telescope (HE, 20-250 keV), Medium Energy telescope (ME, 5-30 keV), and Low Energy telescope (LE, 1-10 keV). Here we present the detailed observation results of all three telescopes of \insight~ on the prompt emission of GRB 221009A. After dead-time and data saturation correction, we recovered the light curves of HE, ME and LE telescopes and find that they generally track the GECAM-C low gain light curves that are free of data saturation issues.
Particularly, the ME light curve matches the GECAM-C light curve in low gain mode above 400 keV, while the LE light curve is more consistent with the GECAM-C above 1.5 MeV.
Based on simulation, we find that the signals recorded by the ME and LE are actually caused by the secondary particles produced by the interaction between GRB gamma-ray photons and the material of the satellite.
Interestingly, the consistency between ME and LE light curves and GECAM-C demonstrates that ME and LE data could be used to characterize the GRB properties. 
Espeically, the high time resolution light curve of ME  allowed us, for the first time, to calculate the minimum variability timescale (MVT = 0.10 s) of the main burst episode of GRB 221009A.

\end{abstract}


\section{Introduction} \label{sec:intro}

GRB 221009A, also known as Swift J1913.1+1946, is an exceptionally bright and long-duration gamma-ray burst (GRB) discovered on October 9, 2022. This event was jointly detected by many ground- and space-based telescopes, from radio to gamma-ray \citep[e.g.][]{2022ATel15660....1T,2022GCN.32636....1V,2022GCN.32632....1D,2022GCN.32648....1D,2022GCN.32668....1F,2022GCN.32650....1U,2022GCN.32660....1G,2022GCN.32677....1H,gecam09A2023arXiv230301203A,2023ApJ...946L..21N,2023A&A...677L...2R,2023ApJ...949L...7F,2024NatAs...8..774B,LAT_09A,2023ApJ...946L..24W,Radio_09A_24}, providing a unique opportunity to study the GRB physics in unprecedented detail. Remarkably, GRB 221009A is one of the closest gamma-ray bursts till today, with a redshift of $z=0.151$ \citep{2022GCN.32648....1D}, and among the brightest bursts ever detected. Indeed, the conclusion that it has the highest flux and fluence was made even from a preliminary estimation of the saturated data \citep[e.g.][]{GBM_09A}, which was confirmed by the accurate measurement by GECAM-C (the third member of Gravitational wave high-energy Electromagnetic Counterpart All-sky Monitor (GECAM), also
known as High Energy Burst Searcher, HEBS) \citep{gecam09A2023arXiv230301203A}.  

Regarding the prompt emission, the exceptional brightness prevented many gamma-ray telescopes (e.g. Fermi/GBM and \insight/HE) from making accurate measurements of the main burst because of instrumental limitation (e.g. data saturation and pulse pileup). With its unsaturated data especially during the bright episode, GECAM-C provided an accurate and high resolution measurement of the light curve and spectrum of the prompt emission, which consists of a precursor, two bright bumps in the main burst region, and a bight flare. The spectral analysis of GECAM-C data shows that GRB 221009A has a record-breaking isotropic energy $E_{\rm iso}\sim1.5\times10^{55} \rm erg$ \citep{gecam09A2023arXiv230301203A}.
The occurrence rate of such a bright event is estimated to be approximately once every 10,000 years or slightly higher \citep{boat2023ApJ...946L..31B,Malesani2003}, dubbed Brightest Of All Time (BOAT) GRB.

Regarding the afterglow emission, although the exceptional brightness of GRB 221009A saturated the \textit{Fermi}/LAT
\citep{2022GCN.32637....1B,2022GCN.32658....1P,2022GCN.32760....1O,2022GCN.32916....1O,LAT_09A}, the Water Cherenkov Detector Array (WCDA) of Large High Altitude Air Shower Observatory (LHAASO) recorded more than 64,000 high-energy photons $>$0.2 TeV from the burst, the first time of detailed observations of GRB afterglow onsets \citep{2022GCN.32677....1H,2023Sci...380.1390L}. Moreover, the Kilometer Squared Array (KM2A) of LHAASO detected some photons up to about 13 TeV \citep{2023SciA....9J2778C}. The TeV detection has made important impacts not only on the GRB study but also on the fundamental physics
\citep[e.g.][]{2024PhRvL.133g1501C,zhang2024TeVSOC}.
The X-ray afterglow is hundreds of times brighter than that seen before, making GRB 221009A the seventh gamma-ray burst with X-ray rings \citep[e.g.][]{2023ApJ...946L..30T,2023MNRAS.526.2605S,2023ApJ...946L..24W}. The optical and radio afterglow has been monitored extensively and showed complicated features \citep[e.g.][]{2023GCN.33243....1G,2023ApJ...946L..23L}. Importantly, an achromatic jet break feature was clearly identified in the early afterglow from the keV to MeV band \citep{2024ApJ...962L...2Z} up to the TeV band \citep{2023Sci...380.1390L}.
In addition, a close relation between the prompt emission and TeV afterglow has been found, supporting the scenario that the jet injected energy into the external shock continuously \citep{2024ApJ...972L..25Z}.

Exceptionally, profound spectral line features have been reported by two separate studies. Only using the Fermi/GBM data, \cite{ravasio2023brightmegaelectronvoltemissionline} searched the burst excluding most of the main burst region where the Fermi/GBM data suffered instrumental limitation and reported the emission line features around 10 MeV with a nearly constant line width. Meanwhile, with the joint observation data of GECAM-C and Fermi/GBM, \cite{2024SCPMA..6789511Z} searched the full burst including the brightest region and found that the emission line actually evolves from 37 to 6 MeV. Both the central energy and flux of the line decrease with time as power-law functions, while the line width to central energy ratio remains nearly constant, at about $10\%$. These characteristics of the emission line play a crucial role in unraveling the jet physics and emission mechanisms of GRBs \citep[e.g.,][]{2024SCPMA..6789511Z,2024Sci...385..452R,2024ApJ...968L...5W,2024ApJ...973L..17Z,2024arXiv240716241P,2024arXiv240908485Y}.

On the other hand, the extremely high flux of the prompt emission of GRB 221009A also induced some interesting and rare phenomena. It caused mild perturbations in Earth's ionosphere for several hours \citep{2022RNAAS...6..222H,2023Atmos..14..217P,Cheng_2024}. Gamma-rays from the GRB also produce secondary particles when they interact with the materials of spacecraft and instruments. \cite{battison2023ApJ...946L..29B} conducted an investigation into the source of the anomalous flux detected by the HEPP-L and found that these fluxes perfectly align with the GRB time history. Monte Carlo (MC) simulation
supports the interpretation that high-energy gamma rays from the GRB interact with the detector structure close to the silicon sensors, leading to the production of prompt electrons that partially deposit their energy in the sensors.
\cite{agapitov2023ApJ...948L..21A} also reported the effects of GRB 221009A by the spacecraft particles detectors on board the probes of the THEMIS and ARTEMIS project. 

Considering all these facts, GRB 221009A is indeed a very rare, interesting and important event. In this work, we present the results of the \insight~observation of GRB 221009A with its three telescopes, with focus on the temporal behavior of the main burst episode which is inaccessible for many other instruments. We will begin with an overview of the observation facts of \insight and describe our data analysis methods employed in the study. Subsequently, we will present the simulation results regarding the observation interpretation. Finally, we will discuss the results of data analysis and provide a summary.

\section{Observation and data reduction} \label{section2}

\begin{figure}[http]
\centering
\includegraphics[width=\columnwidth]{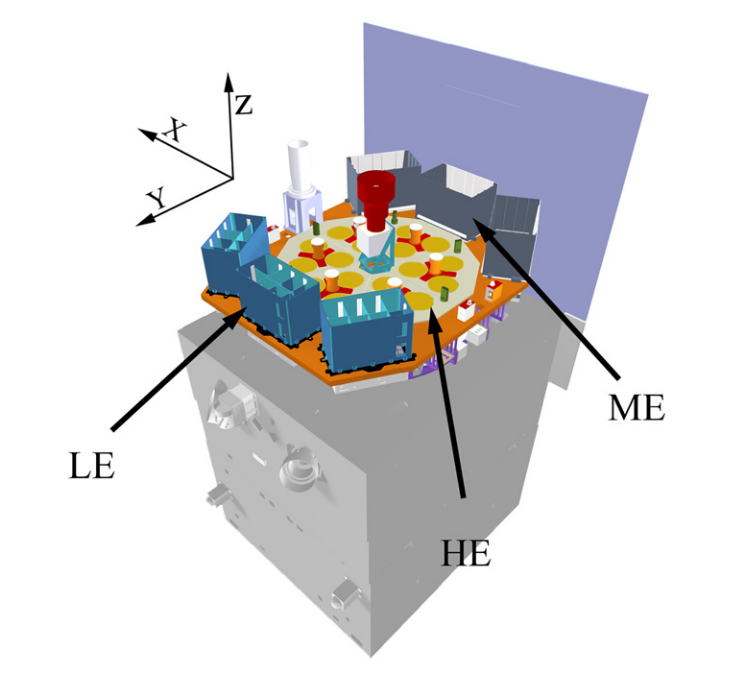}
\caption{The illustration of \insight~ mass model with satellite platform and the payloads of HE, ME and LE telescopes. The coordinates is shown on the upper left.}
\label{HXMTMassModel}
\end{figure}

\textit{Insight}-Hard X-ray Modulation Telescope (\textit{Insight}-HXMT) is China's first X-ray astronomy satellite \citep{2020SCPMA..6349502Z,2020JHEAp..27...64L} which was launched on June 15th, 2017 to the orbit of altitude of 550 km and 
inclination of 43 degree. As shown in Figure \ref{HXMTMassModel}, \textit{Insight}-HXMT features a broad X-ray range (1-250 keV) composed with three telescopes: the High Energy X-ray telescope (HE) which mainly uses 18 NaI(Tl)/CsI(Na) phoswich scintillation detectors for 20-250 keV \citep{2020SCPMA..6349503L}, 
the Medium Energy X-ray telescope (ME) which uses 1728 Si-PIN detectors for 5-30 keV \citep{2020SCPMA..6349504C}, and the Low Energy X-ray telescope (LE) which uses 96 Swept Charge Device (SCD) detectors for 1-15 keV \citep{2020SCPMA..6349505C}. It is worth to note that, apart from the NaI/CsI detector, HE is also equipped with the Anti-Coincidence Detectors (ACDs) and Particle Monitors (PMs) \citep{2020SCPMA..6349503L}.

All three telescopes of \insight~make use of slat collimators to confine their field of views in the normal observations, either pointed observation or scan observation, of various X-ray sources \citep[e.g.][]{2021NatAs...5...94M,2021NatAs...5..378L,2023Sci...381..961Y}. 

On contrast, since the high energy gamma-rays can penetrate the satellite structure and leave signal in the CsI detectors of HE, the CsI can effectively monitor the all-sky in gamma-ray band ($>$ about 200 keV), especially for GRBs \citep{2022ApJS..259...46S,2023MNRAS.518.2005C}. Indeed, HE also participated in the monitoring of the first GW electromagnetic counterpart and provided stringent constraints in MeV energy band \citep{Li:2017iup}. 
Detailed studies show \textit{Insight}-HXMT works in a good status in terms of temporal and spectral observations \citep{2023arXiv230210714L,2023arXiv230203859L,2023arXiv230214459L}, including the GRB observation \citep{LUO20201}.

\subsection{\textit{Insight}-HXMT~observation}
 
\begin{figure}[http]
\centering
\includegraphics[width=\columnwidth]{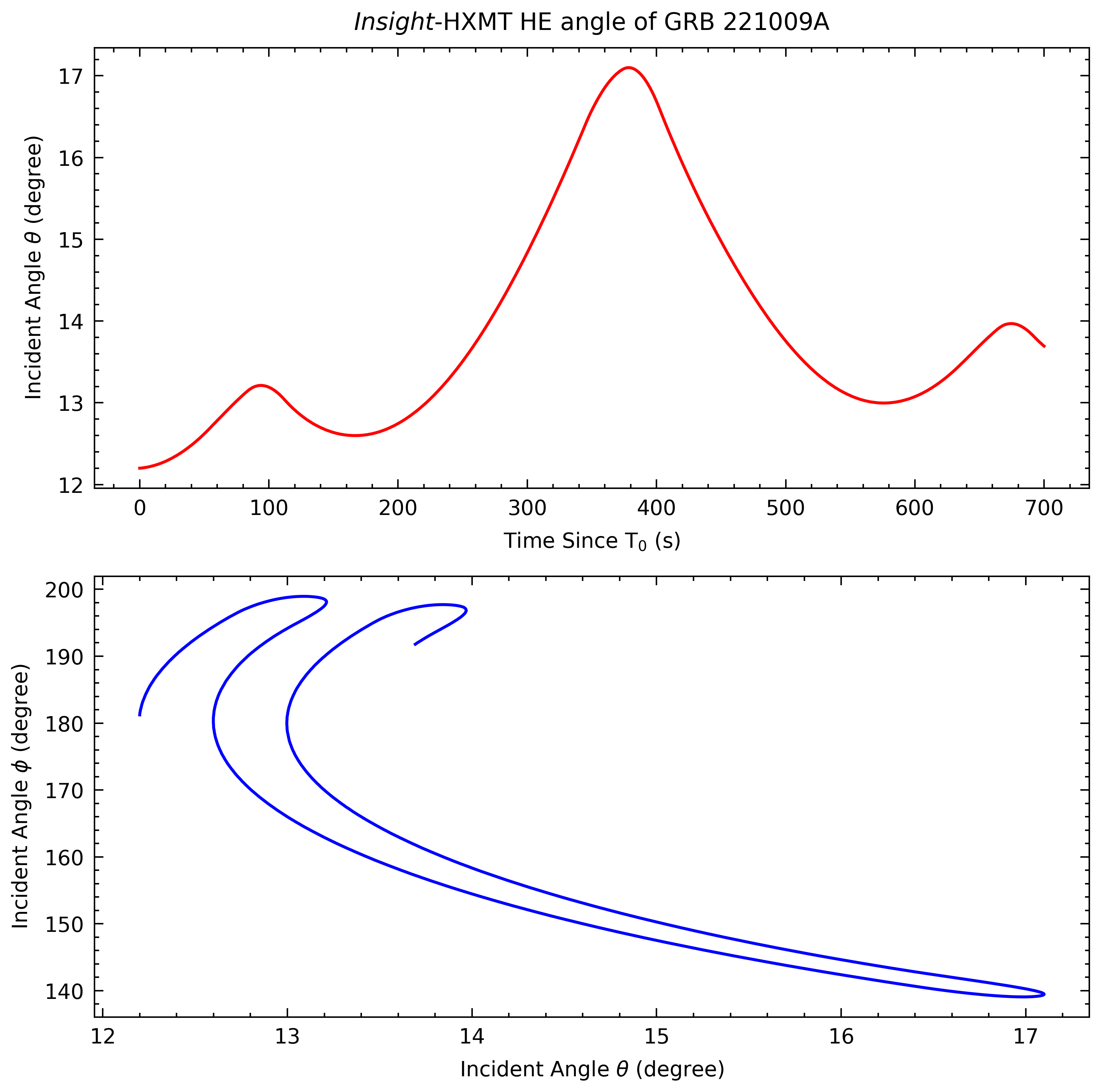}
\caption{Incident angles of GRB 221009A with respect to \insight. \insight~is executing Galactic Plane Scanning Survey. Top panel: the incident angle $\theta$ changes with the time since trigger time $T_0$. Bottom panel: the incident angles $\theta$ and $\phi$ during $T_0$ to $T_0$+700\,s. The time evolution could be inferred from the top panel.}
\label{Mutipl_sat_angle}
\end{figure}

\textit{Insight}-HXMT was triggered by the precursor of GRB 221009A at 13:17:00.050 UT on October 9th, 2022 (denoted as $T_0$) during a routine ground search for bursts \citep{2022ATel15660....1T} with the CsI(Na) detectors of the High Energy X-ray telescope (denoted as HE/CsI hereafter), which has regularly detected GRBs \citep{2021MNRAS.508.3910C,2022ApJS..259...46S}. 
The incident angle of GRB 221009A was shown in Fig.\ref{Mutipl_sat_angle}. Notice that, when this burst was detected, the \textit{Insight}-HXMT was executing Galactic Plane Scanning Survey and the position of this GRB is fortunately quite close to the boresight of telescopes, which is beneficial for the detection.

When GRB 221009A occurred, the HE/CsI detectors of \textit{Insight}-HXMT operated in the low voltage mode (PMT voltage lower than normal value) with the energy range of about 200-3000 keV (deposited energy). Only gamma-rays with energy greater than about 200 keV can penetrate the spacecraft and leave signals in the HE/CsI detectors installed inside of the telescope. 

 From the precursor to the main emission and the flare, as well as the early afterglow phase, GRB 221009A was visible by \insight ~until it entered the Earth's shadow at $T_0$+1884 s. 
During the full burst of GRB 221009A, \insight ~was making scan observations to a small sky area, and its pointing direction was moving (see Figure \ref{Mutipl_sat_angle}), so the incident angle of GRB 221009A to the CsI detectors was changing. For the time-resolved spectral analysis, we make the CsI energy response matrix according to the incident angle averaged in the time interval.

The main emission and the flare peak are so bright that ACDs and PMs of HE as well as ME and LE also recorded the GRB signal despite that their detection efficiency to gamma-rays is very low. Data saturation occurred in HE/CsI detectors during some time period with ultra high count rate (See \cite{2020JHEAp..26...58X} for details), while PMs are free of such saturation because its count rates is not extremely high due to the very small detector areas and the aforementioned low efficiency to gamma-rays. However, the relatively higher rates of PMs than normal case still triggered the threshold of turning off all other detectors of HE and ME from about $T_0$+230\,s to $T_0$+250\,s and $T_0$+268\,s to $T_0$+278\,s, just like the satellite was passing the South Atlantic Anomaly (SAA) region. 
Fortunately, LE did not shut down but entered a special working mode during the SAA passage. This feature allows us to recover the LE data to obtain useful information of this burst, especially the bright emission region of the burst. 

\subsection{Data and Methods}\label{section2.2}
The extremely high flux of the burst resulted in the detected events surpassing the storage limits of the on-board data buffer, causing the observed data to saturate. This saturation means that some events are lost during certain time intervals. In addition to saturation, the deadtime effect is also getting into play. The deadtime occurs when the front-end electronics of a detector process an event, preventing all six detectors sharing the same Physical Data Acquisition Unit (PDAU) with this detector from recording new photons \citep{2020JHEAp..26...58X}. Both HE and LE were severely affected by data saturation, whereas HE and ME were significantly influenced by deadtime. Corrections for both effects are necessary.

\begin{figure}[http]
\centering
\includegraphics[width=\columnwidth]{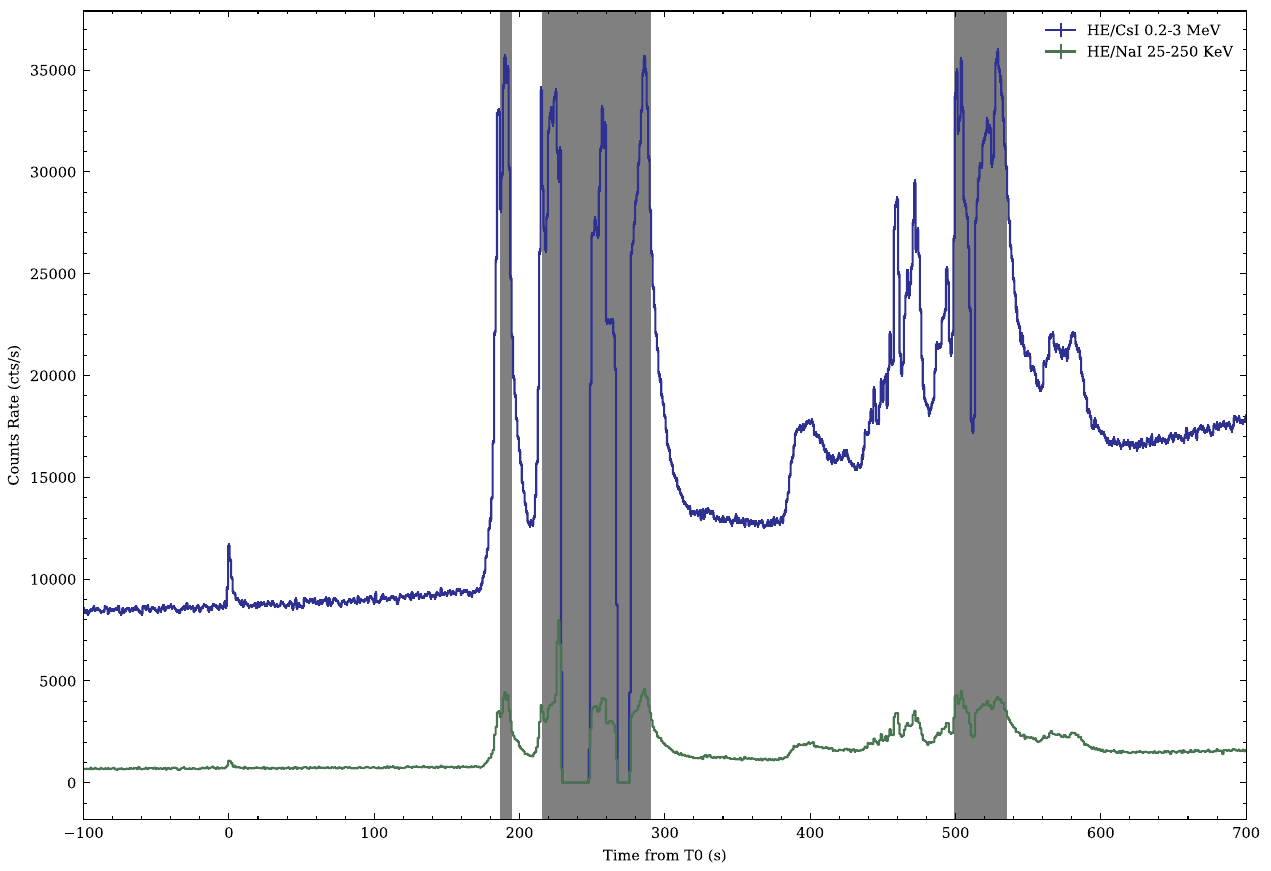}
\caption{HE raw light curves. The blue line shows the CsI raw light curve during $T_0$-100 s to $T_0$+700 s. The green line presents the NaI raw light curve. The grey regions indicate the time intervals of data saturation and the SAA region working mode.}
\label{he_lc}
\end{figure}

HE clearly experienced major data saturation effects, especially during the main emission period from approximately $T_0$+185\,s to $T_0$+290 s. Although saturation corrections usually could be applied to HE \citep{2023ApJ...953...67G}, it is hardly to do this correction for this exceptionally bright GRB due to a very high portion of data saturation during the main burst episode.
Figure \ref{he_lc} illustrates the raw light curves for the CsI and NaI detectors of HE.
We can see that the HE data is heavily affected by saturation loss during the main emission phase, therefore we will not analyze the HE data in more detail but focus on the ME and LE data hereafter.

\begin{figure}[http]
\centering
\includegraphics[width=\columnwidth]{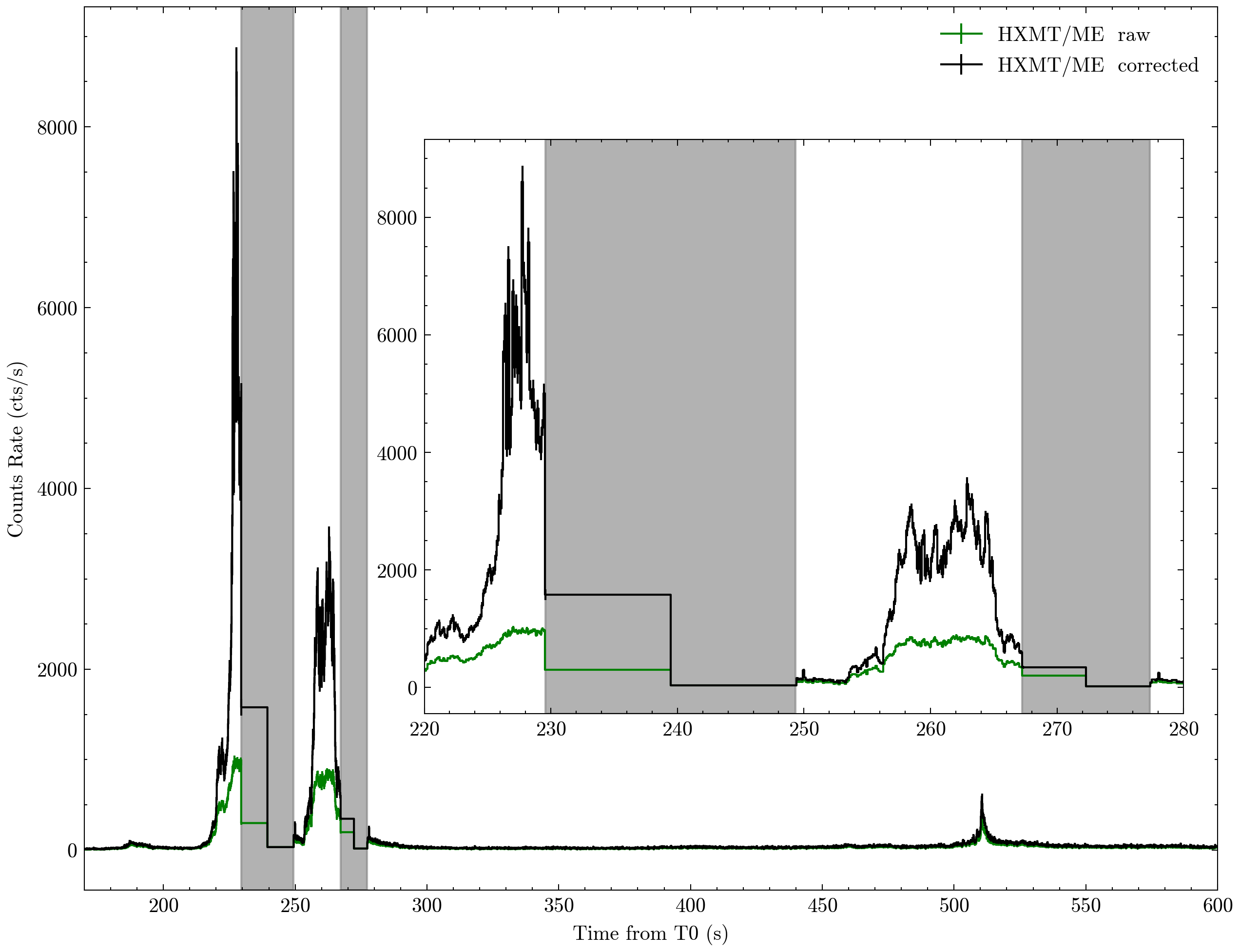}
\caption{ME light curves. The black line shows the dead-time corrected ME light curve during $T_0$+170 s to $T_0$+600 s. The green line presents the ME raw light curve. The grey regions are the time intervals of the SAA region working mode.}
\label{me_lc}
\end{figure}

ME was unaffected by data saturation but there were gaps in the raw data. The HXMTDAS \citep{2020ASPC..527..469Z} can calculate ME's deadtime, with Figure \ref{me_lc} showing ME lightcurves before and after deadtime corrections.
Unfortunately, from about $T_0$+230 s to $T_0$+250 s and $T_0$+268 s to $T_0$+278 s, during the period when \insight ~was operating in the SAA working mode, ME was shut down and thus no data was produced for these time intervals, as indicated by the gray shaded region in Figure \ref{me_lc}. However, it is important to note that the ME data covered a good part of the first bump and nearly the complete second bump during the main burst episode of GRB 221009A.

The situation of LE is quite different from ME and HE. LE is composed of three detector boxes, each with 32 SCD detectors. LE can process the data from each detector box independently. In addition to standard physical events with energies surpassing the on-board threshold, LE also records forced trigger events. These events log the amplitude of the noise or the pedestal offset for each SCD detector every 32ms \citep{li2020JHEAp..27...64L}. If there is no saturation effect, the count rate of the forced trigger events in each detector box would be 1000 counts per second. Although LE experienced saturation during two intervals (from $T_0$+225 s to $T_0$+230 s and from $T_0$+257 s to $T_0$+266 s), the recorded forced trigger events' count rate could be used to correct the saturation in the LE light curves, considering that the three detector boxes have different saturated time intervals.

During two periods, approximately from $T_0$+230 s to $T_0$+250 s and $T_0$+268 s to $T_0$+278 s, LE entered the SAA region working mode. In this mode, one of the four boards in a single LE detector box would collect a forced trigger event every 1 ms, with a 10 us collection time window. This allowed the other three boards in the same detector box to collect normal event data within this window, enabling signal collection under the SAA region mode.

Following the SAA region working mode analysis, we multiplied the collected count rate by 100 (derived from 1 ms/10 us) and then by three-fourths to restore the actual signal count rate. This method enabled us to recover the real count rates during these periods, though the margin of error is significant due to the small number of counts actually detected.

\begin{figure}[http]
\centering
\includegraphics[width=\columnwidth]{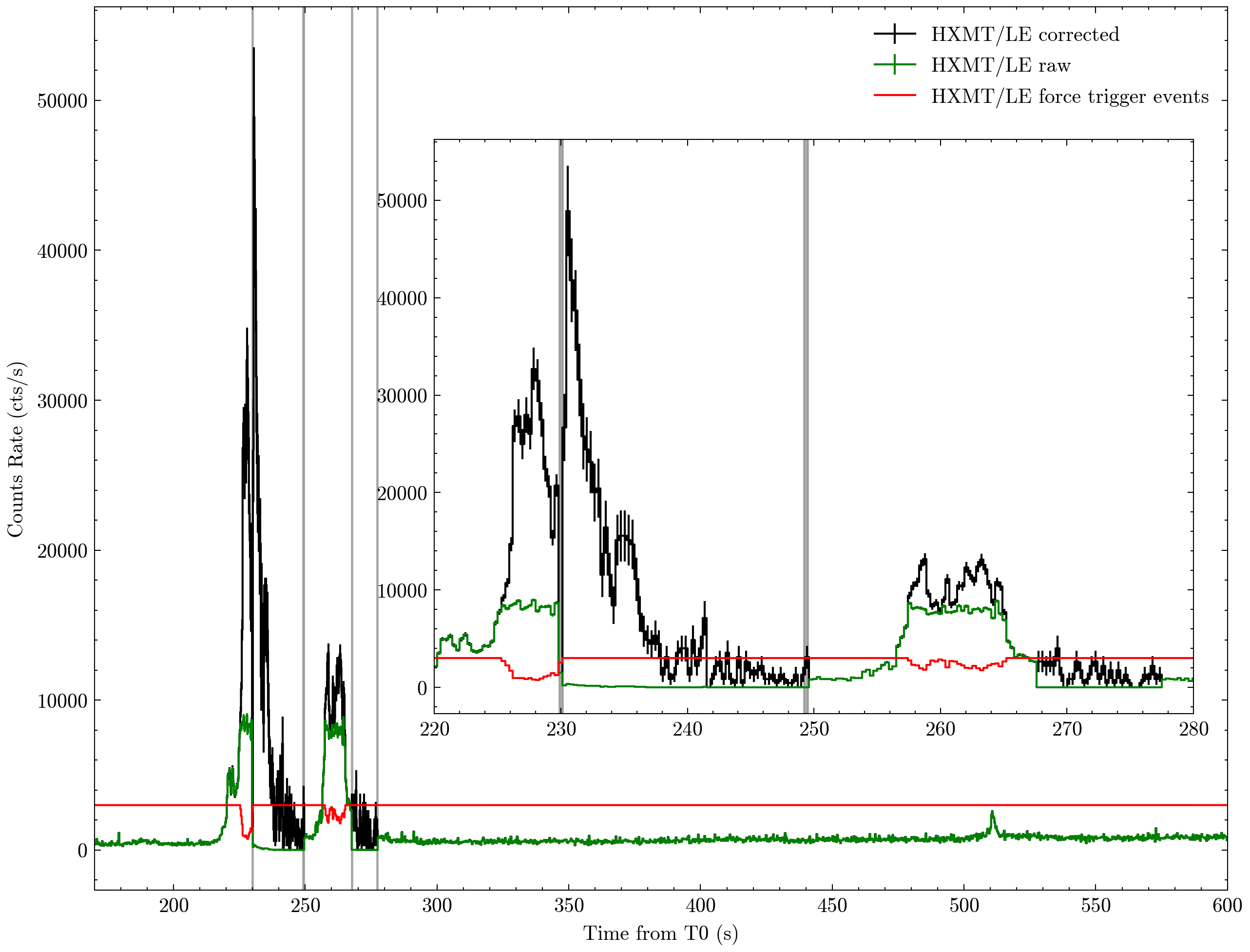}
\caption{LE light curves. The black line represents the LE corrected light curve with a bin size of 0.3 s, the green line represents the LE raw light curve, and the red line represents the LE light curve of the force trigger events.
The gray regions show the time when LE entered and exited the SAA region working mode, the LE light curve's correction factor may require further refinement in the regions.}
\label{le_lc}
\end{figure}

Finally, we managed to reconstruct the LE light curve with nearly full-time coverage of this burst, which is very helpful to study this burst. Figure \ref{le_lc} presents the LE light curves before and after the saturation and SAA region working mode corrections. We note that the dead time of LE caused by the forced trigger events is a minor and negligible issue \citep{2020SCPMA..6349505C}.

\section{Simulation} \label{section3}
Because this burst is out of the field of view of ME and LE, and their detection efficiency of gamma-rays is very low, it was initially unclear to us why ME and LE recorded signals from this burst. To understand the light cures of LE and ME, we first checked the ME and LE detector responses to gamma-ray photons with a series of dedicated Monte Carlo simulations, which are performed using the Geant4 toolkit.
A mass model of \insight~ payload and satellite platform is constructed, as shown in Figure \ref{HXMTMassModel}. In order to include the process of photo-electron, scattering, photon conversion and other interactions happening between X-ray and the materials, a computer-aided design (CAD) model of a properly simplified \footnote{The simulation will be extremely slow if using the full model of the whole satellite since it consists of a large amount of parts.} satellite platform and payload are included in the simulations, while the sensitive volume of detector and structures close to the detectors are fully constructed (without simplification) with attempts to accurately characterize the real process of detection. The E-C relationship and energy resolution are also taken into account according to the in-orbit calibration results. Gamma-rays with flat spectrum from 0 to 5 MeV were generated from a disc surface placed right at the incident direction of GRB 221009A in the payload coordinates (Figure \ref{HXMTMassModel}). Energy deposition on ME and LE detectors from all particles, including the primary incident gamma-rays as well as secondary particles (such as electrons), are considered. 

\begin{figure}[http]
\centering
\includegraphics[width=\columnwidth]{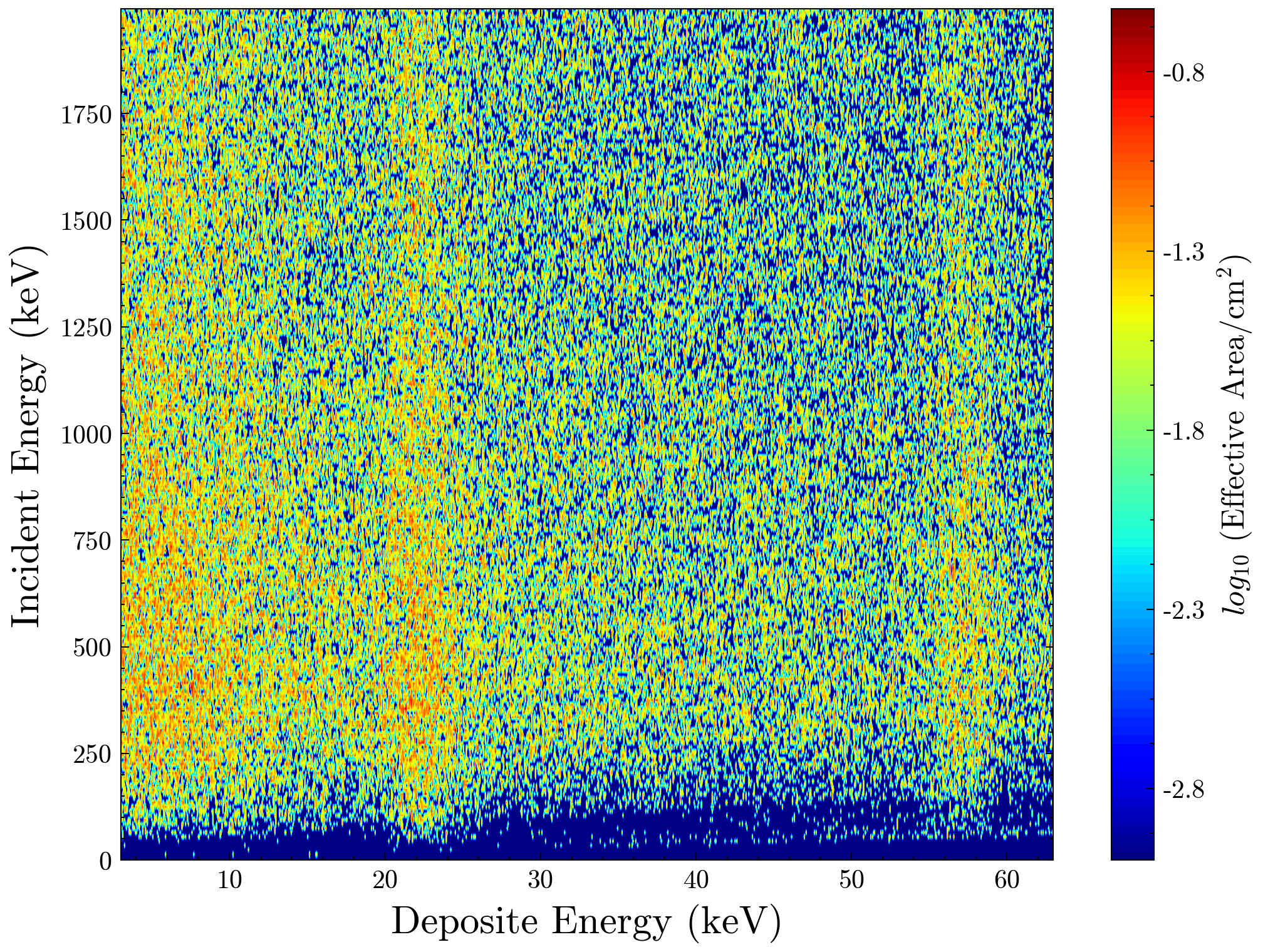}
\caption{Simulated energy response of ME for GRB 221009A.}
\label{MEresponse}
\end{figure}

\begin{figure}[http]
\centering
\includegraphics[width=\columnwidth]{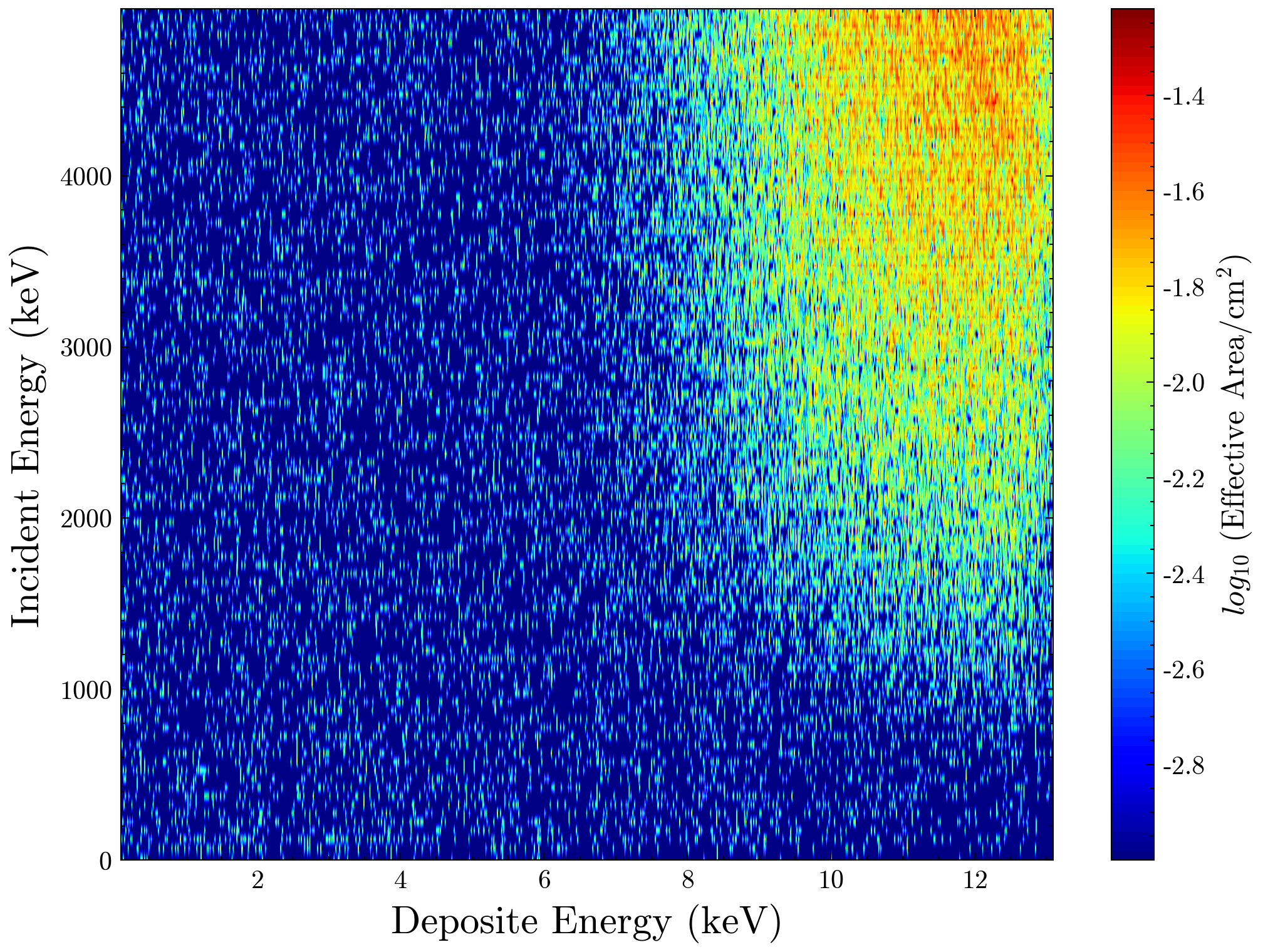}
\caption{Simulated energy response of LE for GRB 221009A.}
\label{LEresponse}
\end{figure}

The response matrix for the LE and ME telescopes are shown in Figure \ref{MEresponse} and Figure \ref{LEresponse}, respectively. Owing to the difference in the design between LE and ME, their detector responses display different characteristics in energy range and redistribution, which mainly come from the thickness of sensitive region, filter, threshold and surrounding shielding material of the detectors. 
Also note that, both LE and ME lost the original energy information of the incident gamma-rays, making it hard to reconstruct the incident energy spectrum.

\begin{figure}[http]
\centering
\includegraphics[width=\columnwidth]{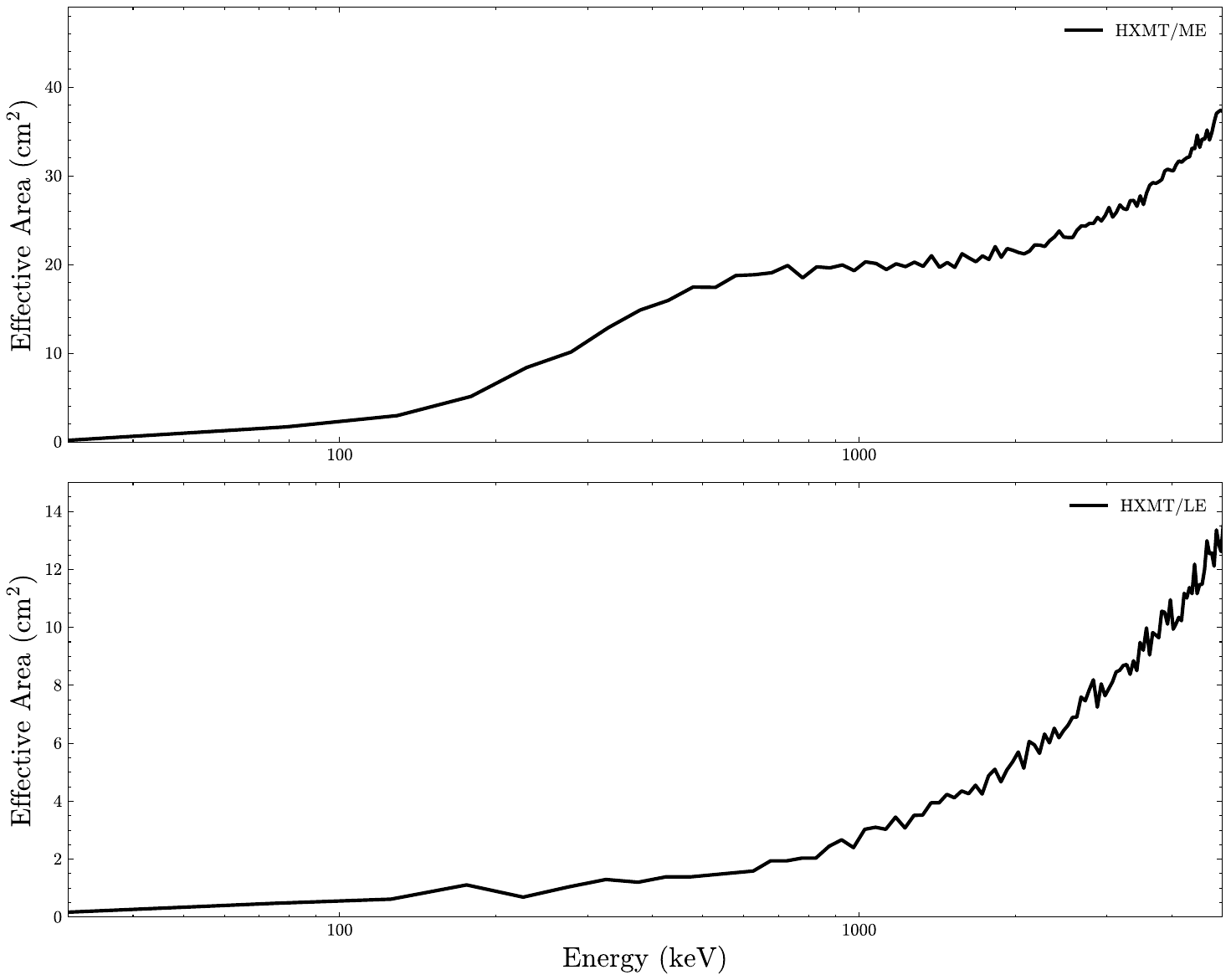}
\caption{Effective area of ME and LE for GRB 221009A derived from the simulation.}
\label{eff}
\end{figure}

As shown in Figure \ref{eff}, the effective area for both LE and ME significantly increases with the energy of the incident photons. Particularly for LE, the increase is notably rapid. For gamma-rays below about 400 keV and 1000 keV, the effective areas for ME and LE are relatively small. Interestingly, this energy range generally corresponds to the energy band of the low gain mode (LG) of GECAM-C.

\section{Results}
\begin{figure*}[http]
\centering
\includegraphics[width=0.7\textwidth, angle=0]{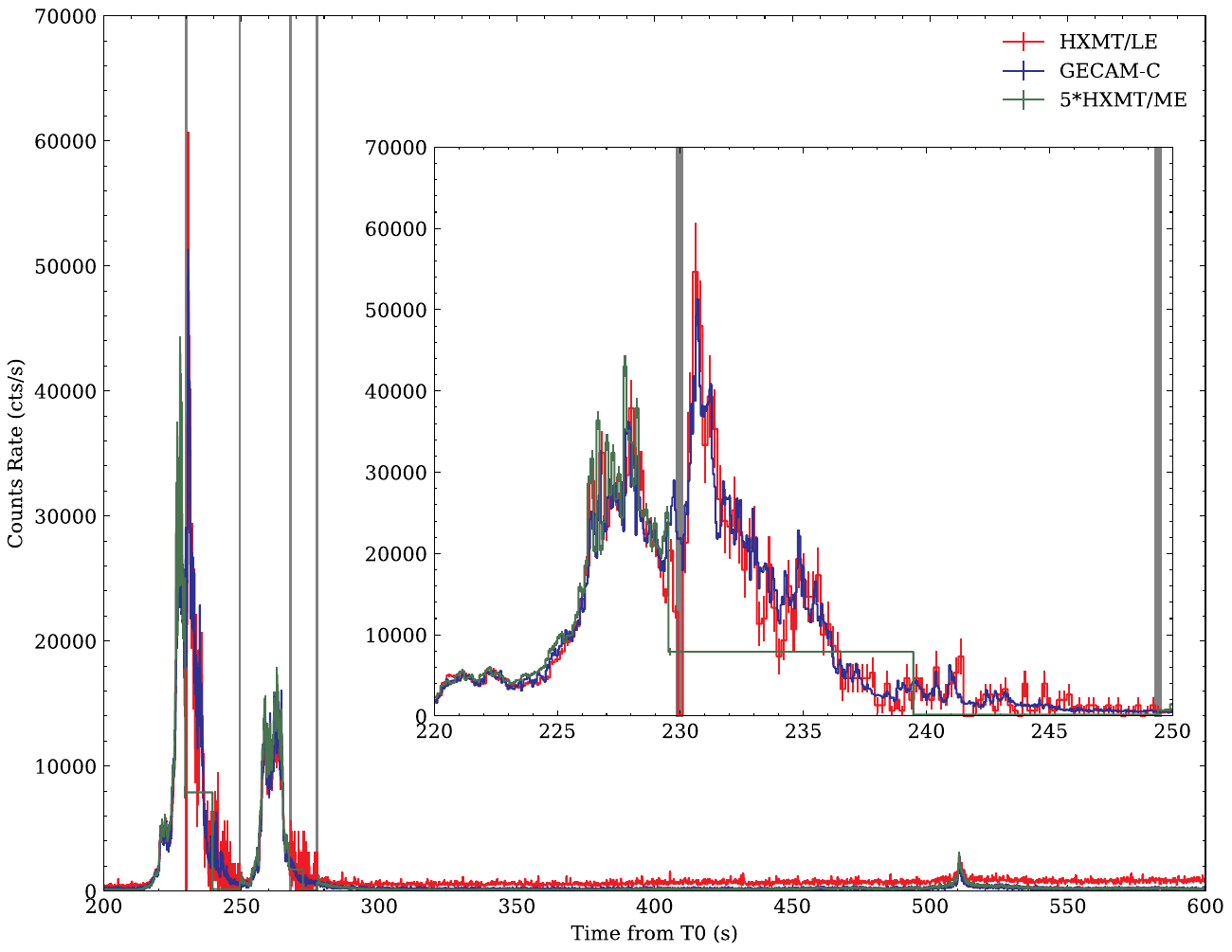}
\caption{Light curve comparison of \insight/ME and LE and GECAM-C low gain. The red line represents the LE  light curve with a bin size of 0.2 s, the green line represents the ME light curve, multiplied by a factor of five, with a bin size of 0.1 s, and the blue line represents the low-gain light curve of GECAM-C with a bin size of 0.05 s.
The gray regions indicate the time when LE entered and exited the SAA region working mode. Note that the correction factor of LE light curve may require further refinement in the regions.}
\label{Mutipl_sat_lc}
\end{figure*}

As mentioned in section \ref{section2.2}, \insight/HE suffered severe and prolonged saturation during the most energetic prompt emission, thus we just used the light curves of ME and LE to compare with the light curve of GECAM-C low gain data, as shown in Figure \ref{Mutipl_sat_lc}. It can be seen that light curves of ME and LE agree with the light curve of GECAM-C very well. 
We also compared their light curves using different time binsizes and found that they match each other under binsize from 0.05 to 1 s.

\subsection{Comparison with GECAM-C light curves}

\begin{figure}[http]
\centering
\includegraphics[width=\columnwidth]{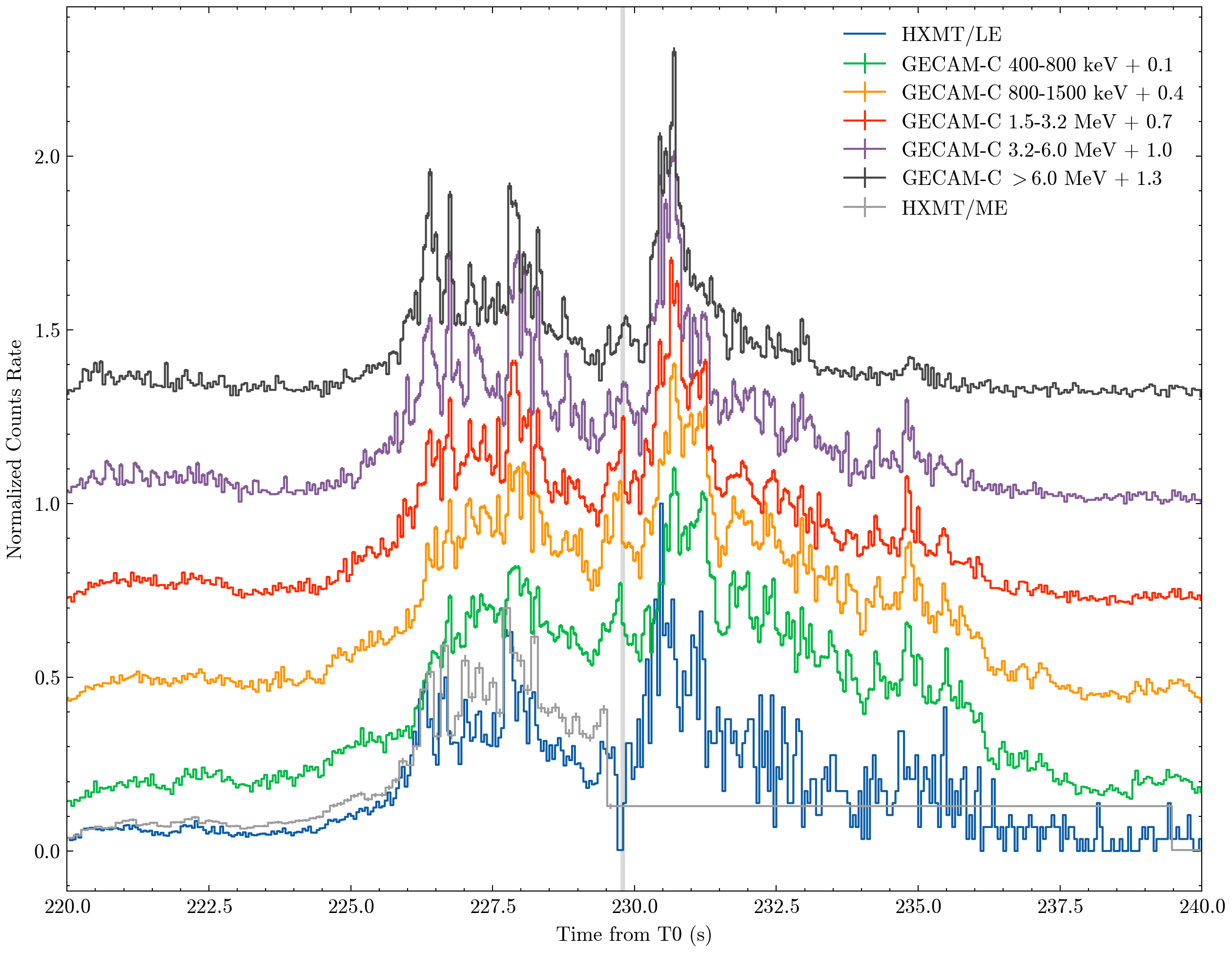}
\caption{Normalized light curves of \insight ~ ME and LE as well as GECAM-C from $T_0$+220 s to $T_0$+240 s in five energy bands.}
\label{5band_lc}
\end{figure}
\begin{figure}[http]
\centering
\includegraphics[width=\columnwidth]{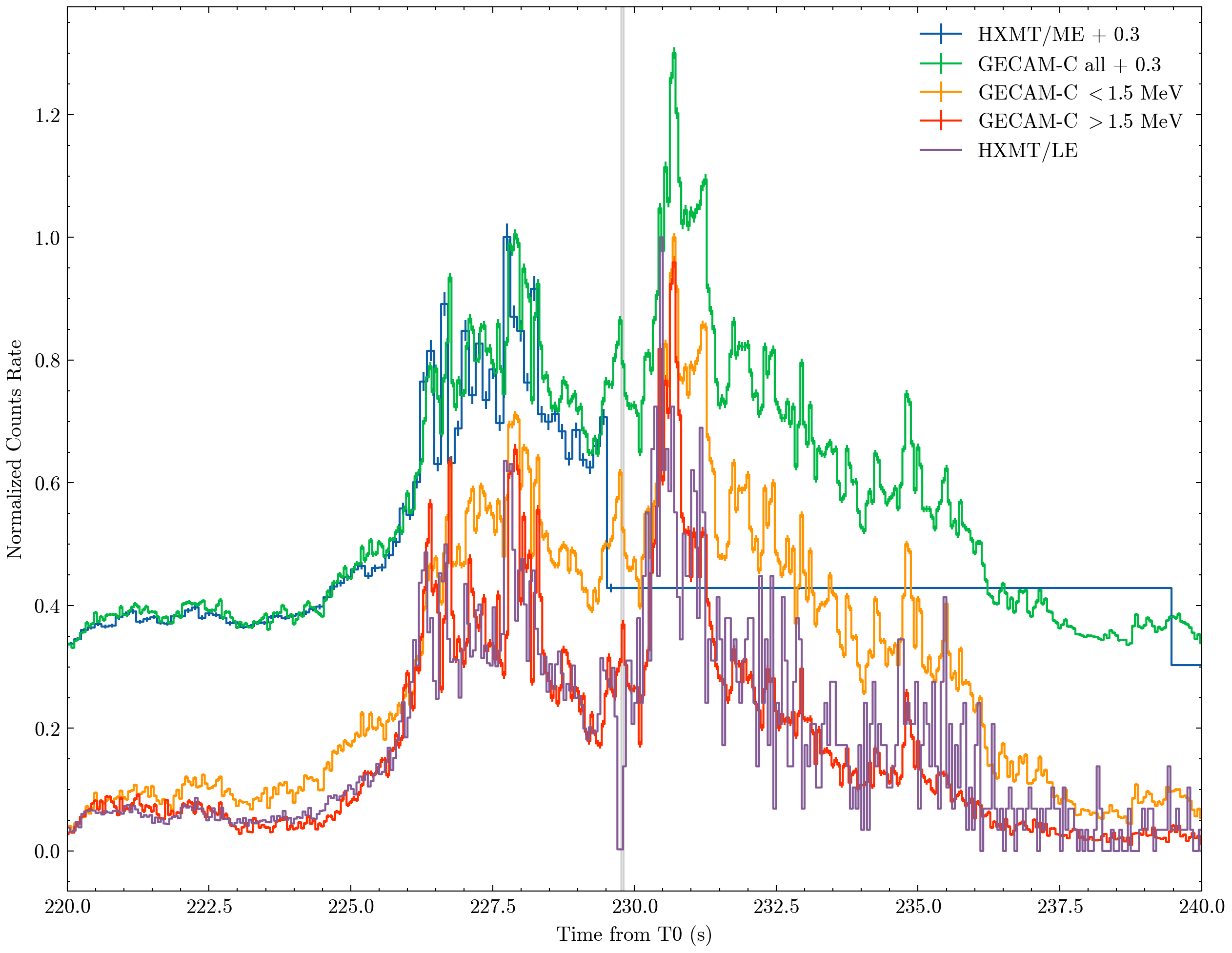}
\caption{\insight ~ME and LE as well as GECAM-C normalized light curves from $T_0$+220 s to $T_0$+240 s in two energy bands.}
\label{Mutipl_lc}
\end{figure}

\begin{figure}[http]
\centering
\includegraphics[width=\columnwidth]{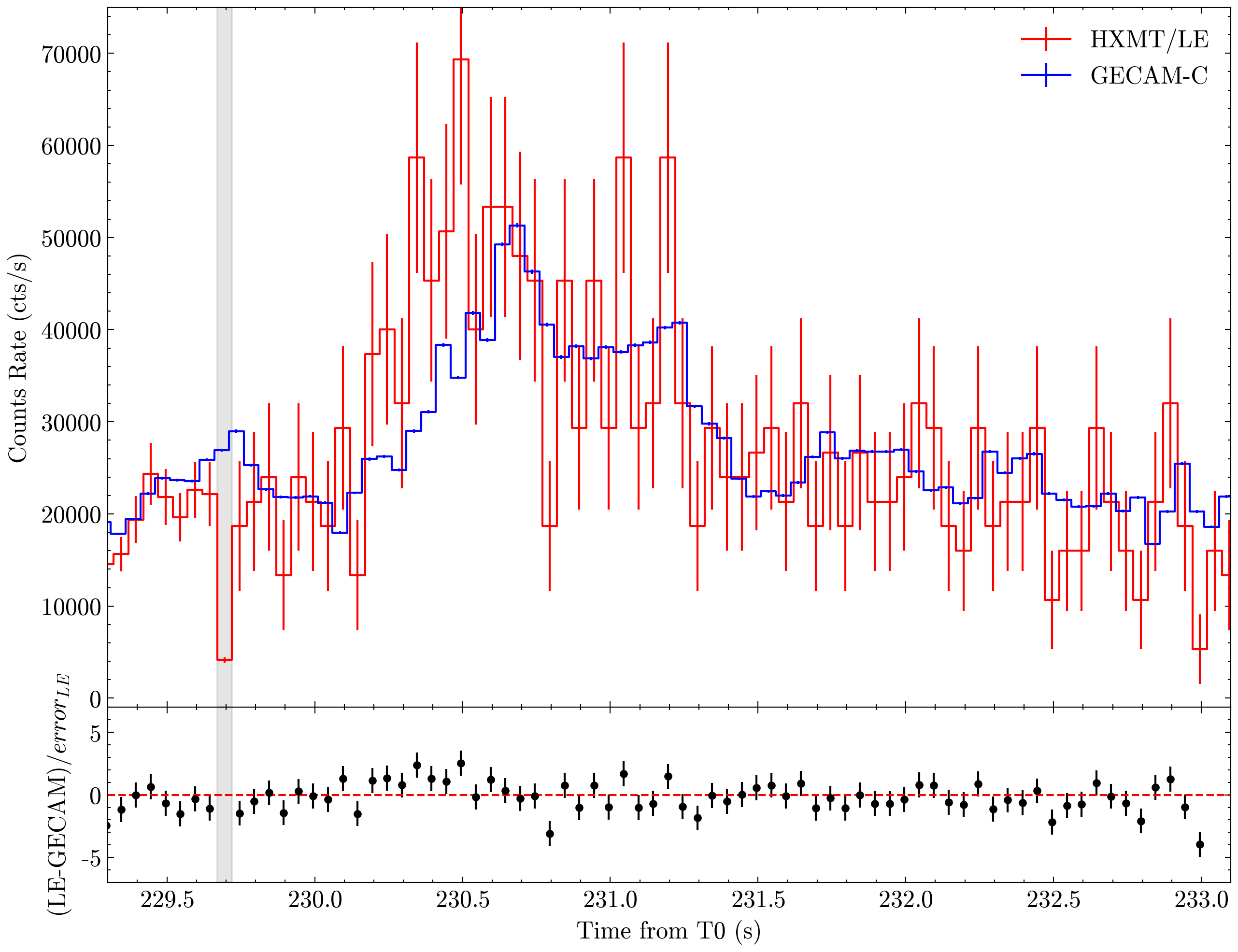}
\caption{Comparison of the light curves of \insight /LE and GECAM-C low gain around the peak of GRB 221009A. The gray regions indicate the time when LE entered and exited the SAA region working mode. Note that the LE light curve's correction factor may require further refinement in the regions.}
\label{peak_compare}
\end{figure}

\begin{figure}[http]
\centering
\includegraphics[width=\columnwidth]{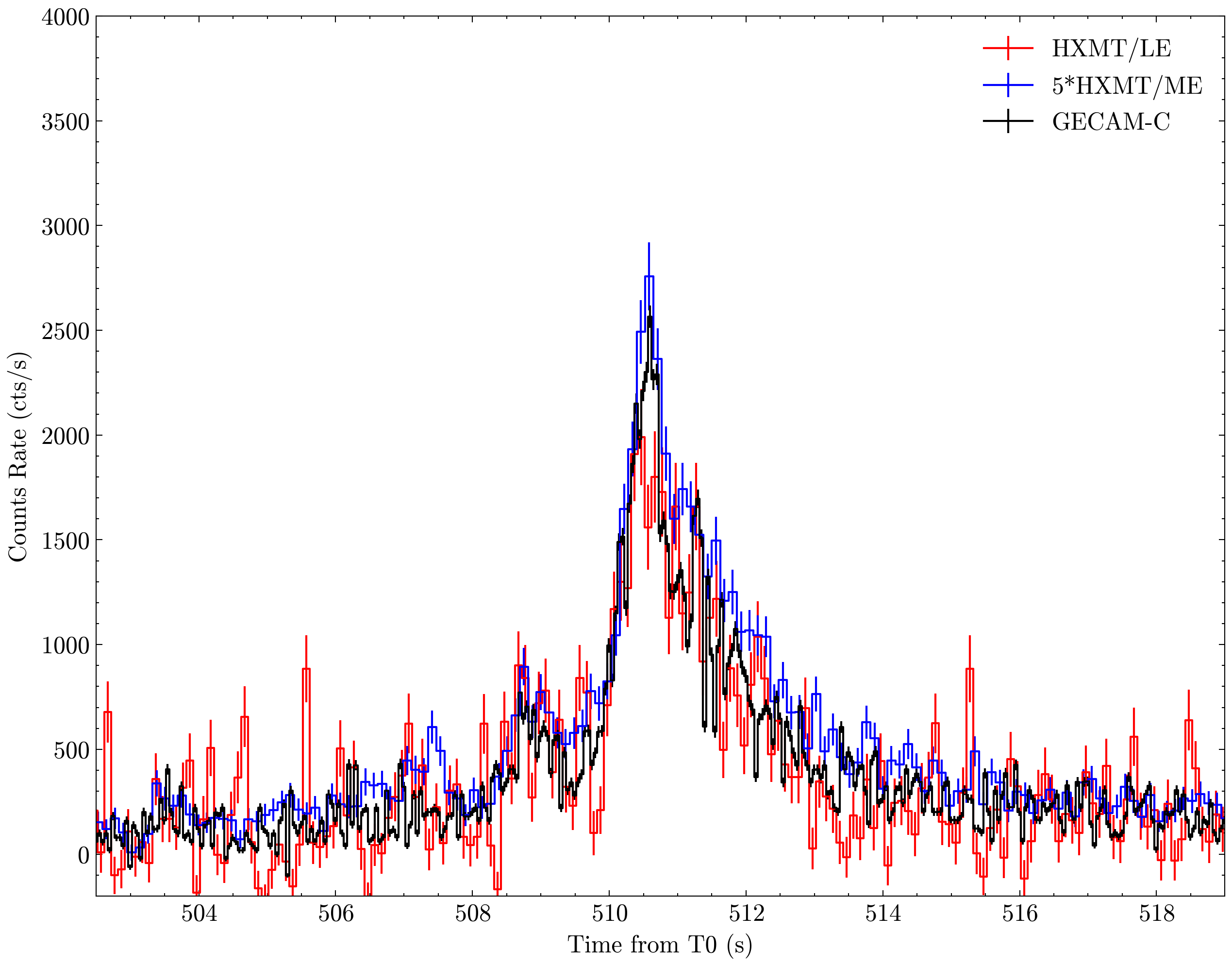}
\caption{Light curves from \insight~ ME and LE as well as GECAM-C low gain during the flare of GRB 221009A.}
\label{flare}
\end{figure}

In order to investigate the correlation between LE, ME, and GECAM-C in detail, we normalized the light curves. Due to the extremely high intensity of some time intervals of this GRB, the background contribution relative to the signal is negligible. Consequently, we chose the approach of maximum normalization, which ensures a uniform scale of amplitudes across all data points.

We further divided the light curve of GECAM-C into five energy bands (Figure \ref{5band_lc}), from which we can see that the light curves in the 400-800 keV and 800-1500 keV bands differ significantly from the light curves in the three energy bands above 1.5 MeV. We found that the main shape of the light curve of LE during the main burst phase of GRB 221009A correlates more with the light curves in the three energy bands above 1.5 MeV of GECAM-C. Therefore, we simply divided the light curves of GECAM-C into two energy bands: above and below 1.5 MeV, and compared them with that of ME and LE, as shown in Figure \ref{Mutipl_lc}. We found that the overall shape of the GECAM-C light curve above 1.5 MeV is more similar to that of LE, while the light curve of ME is most consistent with the GECAM-C light curve in all energy band. 

As presented in section \ref{section3}, we obtained the effective area of incident photons for LE and ME (Figure \ref{eff}). We can see that LE indeed has a larger effective area for incident higher energy photons, especially above 1.5 MeV, while ME has a similar effective area for incident photons above 400 keV. This can partially explain why the light curves of LE and ME agree well with that in different energy bands of GECAM-C. Another possible explanation is that the proportion of LE split events increases with the energy of the incident particles. As a result, higher energy particles are more likely to generate much more split events, leading to the light curve of LE having similar behavior to the GECAM-C light curve above 1.5 MeV.

Besides, we compared the two light curves of GECAM-C and \insight/LE during the most bright part of GRB 221009A, as shown in Figure \ref{peak_compare}. According to the residuals shown in the bottom panel, it can be seen that the light curves of LE and GECAM-C exhibit a high degree of consistency around this bright region where many other instruments suffered sever instrumental effects. 
It is noteworthy that ME and LE also provide observations of the flare of GRB 221009A. Again, their light curve are found to track that of GECAM-C low gain data, as shown in Figure \ref{flare}.

 If any of these corrected light curves of ME, LE and the light curve of GECAM-C low gain experienced significant problem, the consistency between them will not be achieved, therefore, the above finding not only provides an extra support to the conclusion that GECAM low gain data was not affected by instrumental effects, but also indicates that the ME and LE light curves (mostly contributed by secondary particles) can largely reproduce the temporal structure of the original high-energy gamma photons from this GRB. However, since most of the energy information of gamma-rays were lost in the detection process of ME and LE, it is very difficult to obtain the spectral information of this GRB based on ME and LE data.

It is important to note that, in cases of very bright events like GRB 221009A, conventional detectors may face saturation and pulse pileup issues, or similar to GECAM-C, can only capture time-binned data with limited time resolution (i.e. 50 ms for GECAM-C). But the reconstructed light curves of ME and LE in this work do offer data for GRB 221009A, even for the most bright region, with much higher time resolution, i.e. 280~$\mu s$ for ME, and 1 ms for LE \citep{2020SCPMA..6349502Z}, without significant instrumental effects after corrections. Despite that the statistics of the LE light curve is not very high, it provided a very complete time coverage of GRB 221009A with high temporal resolution data.

\subsection{Minimum Variability Timescale}
\begin{figure}[http]
\centering
\includegraphics[width=0.9\columnwidth]{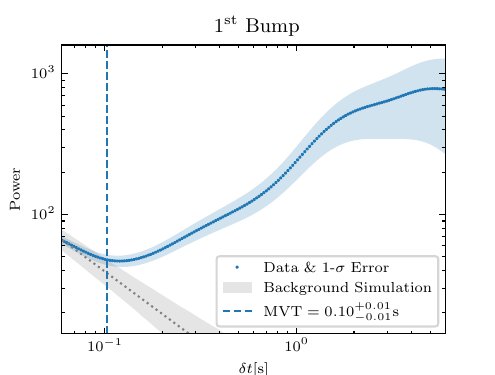}
\includegraphics[width=0.9\columnwidth]{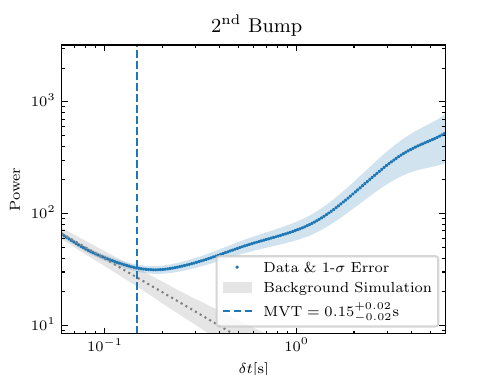}\\
\caption{The MVT results of the first main bump (top) and second main bump (bottom) of GRB 221009A, caculated with the wavelet analysis method \citep{2018ApJ...864..163V}.
}
\label{me_mvt}
\end{figure}

Thanks to its higher statistics, the ME light curve is especially interesting to study. Although the observation of ME was interrupted by the SAA observation mode, the first half part of the first main bump and the full period of the second main bump of the main burst episode of GRB 221009A were fortunately well recorded, enabling high temporal studies for the main burst. With the wavelet analysis method \citep{1998BAMS...79...61T,2018ApJ...864..163V}, we calculated the minimum variability time-scale (MVT) of the two main bumps ($T_0$ + 220 s – $T_0$ + 230 s for the first bump and $T_0$ + 250 s – $T_0$ + 267 s for the second bump, respectively) of GRB 221009A using the dead time-corrected light curve of ME with time bin width of 10 ms. The calculation of MVT is shown in Figure \ref{me_mvt}.
The calculation procedure is as follows: First, generate a set of $10^4$ simulated background light curves, with the same duration and time bin width as the ME light curve after dead time correction. The background rate is estimated in the off-burst interval. Then wavelet spectrum is derived for each simulated background light curve. Subsequently, the 99$\%$ containment interval for each timescale $\delta t$ (gray shaded region) centered on the median value (dotted line) is calculated. The smallest timescale that deviates from the 99$\%$ containment level region of the background is the MVT.

As shown in Figure \ref{me_mvt}, the calculated MVT is 0.10 s$\pm$0.01 s for the first bump and 0.15 s $\pm$0.02 s for the second bump of the main burst of GRB 221009A.
We stress that this is the first time that the MVT of the main burst of GRB 221009A was reported, as many other instruments were unalbe to measure the light curve with sufficient time resolution.

\begin{figure}[http]
\centering
\includegraphics[width=0.4\textwidth]{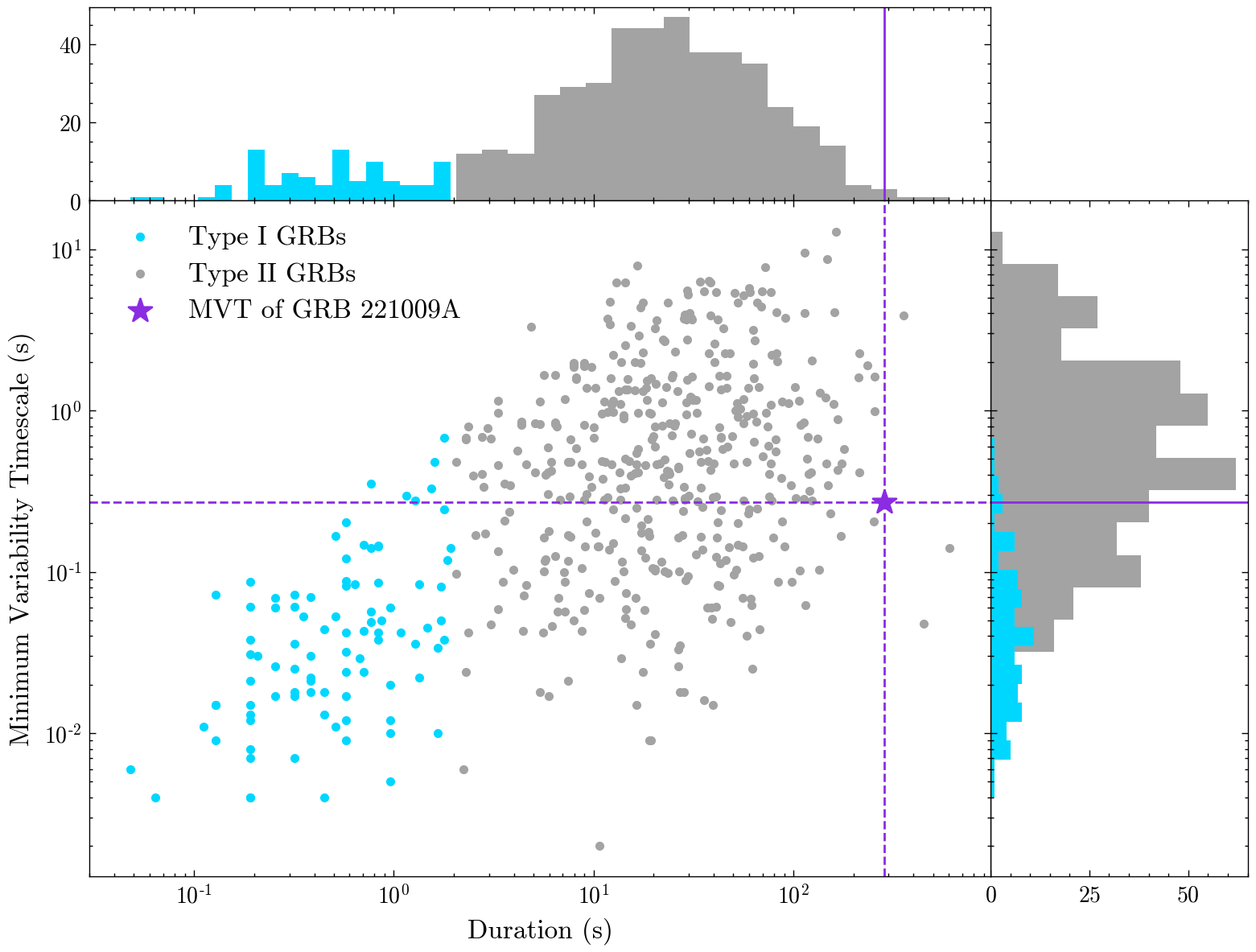}
\caption{The position of GRB 221009A in the minimum variability timescale versus duration diagram. GRB samples are collected from \cite{2015ApJ...811...93G}.}
\label{mvt_t90}
\end{figure}

Although ME did not capture the complete light curve of the first bump of the main burst, the calculated MVT of the first half part of first bump (0.10 s$\pm$0.01 s) is already somewhat smaller than that of the second bump of the main burst (0.15 s $\pm$0.02 s), thus the former (0.10 s$\pm$0.01 s) could be used as the MVT of the main burst.
Since the main burst is the dominant part of this burst, it is reasonable to use the MVT of the main burst to represent the MVT of this burst. Then one can check the the position of GRB 221009A on the MVT-duration diagram, as shown in Figure \ref{mvt_t90}. The duration ($T_{90} $= 285 s) is provided by GECAM-C measurement \citep{gecam09A2023arXiv230301203A}. 
It is evident that, although GRB 221009A is well recognized as long GRB by its relatively long duration, the MVT of GRB 221009A locates in the overlaping region of the long and short GRBs.

Within the internal shock model, the MVT could be used to estimate the radius of prompt emission of the main burst of GRB 221009A: $r_{\rm IS}\simeq 2\Gamma^2ct_{v}\sim6\times10^{15}{\rm cm}(\Gamma/600)^2(t_{v}/0.10~{\rm s})$, where $t_v$ is the variability time derived from ME light curve in this work, and $\Gamma\sim600$ is obtained from the modeling of the relation between keV-MeV and TeV emission \citep{2024ApJ...972L..25Z}. We notice that this estimated radius of the main burst of GRB 221009A is generally consistent with the estimated emission radius (about $10^{16}$ cm) of the MeV emission line which is believed to origin from the main burst episode of the prompt emission \citep{2024ApJ...973L..17Z}.

\section{Discussion and Summary}

In this work, we presented the observation results of all three telescopes of \insight. We reconstructed the light curves for the HE, ME and LE telescopes of \insight~ by applying dead-time and saturation corrections, and found that the ME and LE light curves have a nice agreement with the GECAM-C low gain light curve, which is already proven to be accurate without instrumental effects \citep{gecam09A2023arXiv230301203A}. According to the dedicated simulation for ME and LE, the signal record by them should be the secondary particles produced by the gamma-ray photons from GRB 221009A. By dividing the GECAM-C light curve into two energy ranges, we noted a closer match between the ME light curve and the light curve of GECAM-C above 400 keV, while the LE light curve exhibited a great agreement with the GECAM-C light curve above 1.5 MeV. 
This phenomenon may be attributed to the different energy dependence of effective area for ME and LE, as well as the occurrence of split events of LE.

Since GECAM-C low gain light curves of GRB 221009A are accurate without instrumental effects, the fact that ME and LE light curves tracking the GECAM-C suggest that the secondary particle-induced light curve on ME and LE can reproduce the temporal structure of the gamma-ray light curve of the GRB. Moreover, ME and LE data features a high temporal resolution, which is critically important especially for the study of the main burst region of GRB 221009A, where many other instruments are either suffered by instrumental effects or constrained by low temporal resolution.  

By taking advantage of the high temporal resolution of the ME light curve covering the main burst, we managed to calculate the MVT of the main burst episode of GRB 221009A. We found that, although GRB 221009A generally falls in the long GRB in the MVT-duration diagram, the MVT of GRB 221009A located itself in the overlapping region between long GRBs and short GRBs, and in the lower end of those long GRBs with similar duration as GRB 221009A. This result seems reasonable since the ME data corresponds to the gamma-ray light curve in higher energy band (greater than 400 keV) and the time variation in higher energy is usually faster than the lower energy where the MVT of the GRB sample are usually calculated. 

It is worthy to note that, the MVT (82 ms) of the rising phase before the main burst episode of GRB 221009A is reported based on the Fermi/GBM data \citep{2023ApJ...943L...2L}, however, it is impossible to calculate the MVT of the main burst becuase Fermi/GBM data suffered data saturation effects during the main burst region. As the time scales of the light curve may evlove as the burst goes on \citep[e.g.][]{2007ApJ...667.1024K}, we argue that the measurement of MVT in the rising phase is not necessary equivalent to the main burst region for this extremely bright burst.
Based on our measurement of the MVT of the main burst, we further estimated the radius of the prompt emission region of GRB221009A and find that it is well consistent with other studies.

Lastly, we point out that further studies on the high temporal resolution data of ME and LE may reveal more physics of GRB 221009A, although they did not directly detect the gamma-ray photon of GRB but the secondariy particles produced by the GRB gamma-ray photon.

\section*{Acknowledgments}
This work used data from the Insight-HXMT mission, a project funded by the China National Space Administration (CNSA) and the Chinese Academy of Sciences (CAS). The authors thank supports from National Key R\&D Program of China (Grant No. 2021YFA0718500), 
the Strategic Priority Research Program of Chinese Academy of Sciences (Grant Nos. XDB0550300, 
XDA30050000), 
the Strategic Priority Program on Space Science of Chinese Academy of Sciences (Grant No. E02212A02S), 
the National Natural Science Foundation of China (Grant No. 12273042, 
12373047, 12333007 and 12027803) and International Partnership Program of Chinese Academy of Sciences (Grant No.113111KYSB20190020).

\bibliography{reference.bib}
\bibliographystyle{aasjournal}

\end{document}